\documentclass[aps,prb,twocolumn,tightenlines,floatfix,showpacs,superscriptaddress,10pt]{revtex4-1}
\usepackage{amsmath}
\usepackage{amsfonts}
\usepackage{amssymb}
\usepackage{fancyvrb}
\usepackage{graphicx}
\usepackage{color}
\usepackage{titlesec}
\usepackage{bm}

\global\long\def\avg#1{\langle#1\rangle}

\global\long\def\tr{\mathrm{tr}}

\global\long\def\im{\imath}

\newcommand{\<} {\left\langle}
\renewcommand{\>} {\right\rangle}

\newcommand{\dg} {\dagger}
\newcommand{\pd} {{\phantom\dagger}}

\newcommand{\ci}[1] {c_{#1}^\pd}
\newcommand{\cid}[1] {c_{#1}^\dg}

\newcommand{\ql} {\mathcal{L}}
\newcommand{\qr} {\mathcal{R}}

\newcommand{\qh} {\mathcal{H}}

\newcommand{\bG} {\bm{G}}
\newcommand{\bS} {\bm{\Sigma}}
\newcommand{\bGa} {\bm{\Gamma}}

\renewcommand{\Im}{\operatorname{Im}}

\newcommand{\ue}{\delta \mu}

\titleformat{\section}
{\normalfont\large\bfseries}{\thesection}{1em}{}
\titlespacing*{\section} {0pt}{12pt plus 4pt minus 2pt}{4pt plus 2pt minus 0pt}

\renewcommand{\subsection}[1] {}

\graphicspath{{../}}

\begin{document}

\author{Daniel Gruss}

\affiliation{Center for Nanoscale Science and Technology,
             National Institute of Standards and Technology,
             Gaithersburg, MD 20899}
\affiliation{Maryland Nanocenter, University of Maryland,
             College Park, MD 20742}

\author{Alex Smolyanitsky}

\affiliation{Applied Chemicals and Materials Division,
             National Institute of Standards and Technology,
             Boulder, CO 80305}

\author{Michael Zwolak}

\affiliation{Center for Nanoscale Science and Technology,
             National Institute of Standards and Technology,
             Gaithersburg, MD 20899}

\title{Graphene deflectometry for sensing molecular processes at the nanoscale}


\maketitle

\textbf{Single-molecule sensing is at the core of modern biophysics and nanoscale science, from revolutionizing healthcare through rapid, low-cost 
sequencing~\cite{heerema_graphene_2016,bayley_nanopore_2014,zwolak2008colloquium,branton2008potential,di_ventra_decoding_2016,Kasianowicz1996-1}
to understanding various physical, chemical, and biological processes~\cite{yu_hidden_2017,iversen_ras_2014,gooding_single-molecule_2016,ha_single-molecule_2014,sorgenfrei_label-free_2011,neuman_single-molecule_2008} at their most basic level. However, important processes at the molecular scale are often too fast for the detection bandwidth or otherwise outside the detection sensitivity~\cite{sorgenfrei_label-free_2011,yu_hidden_2017}. Moreover, most envisioned biophysical applications are at room temperature, which further limits detection due to significant thermal noise. Here, we theoretically demonstrate reliable transduction of forces into electronic currents via locally suspended graphene nanoribbons subject to ultra-low flexural deflection. The decay of electronic couplings with distance magnifies the effect of the deflection, giving rise to measurable electronic current changes even in aqueous solution. Due to thermal fluctuations, the characteristic charge carrier transmission peak follows a generalized Voigt profile, behavior which is reflected in the optimized sensor. The intrinsic sensitivity is less than 7 fN/$\sqrt{\mathbf{Hz}}$, allowing for the detection of ultra-weak and fast processes at room temperature.  Graphene deflectometry thus presents new opportunities in the sensing and detection of molecular-scale processes, from ion dynamics to DNA sequencing and protein folding, in their native environment.}

\begin{figure*}
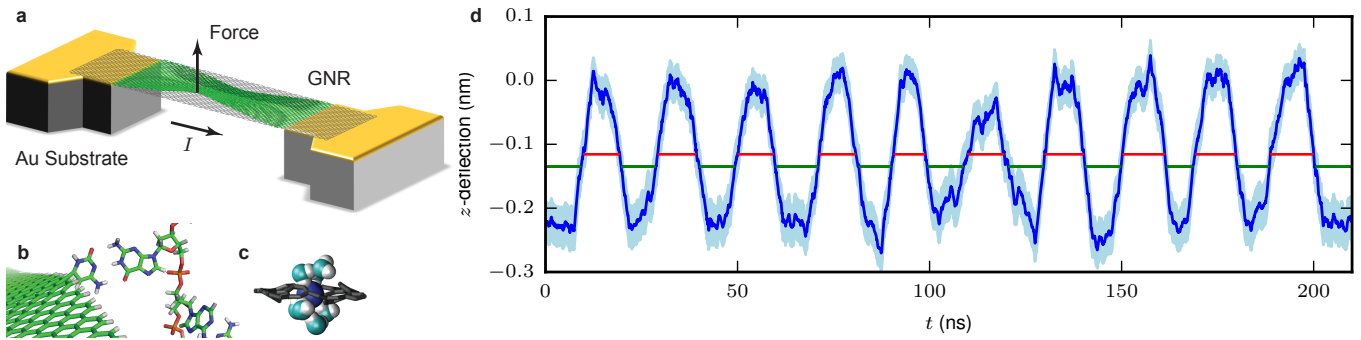

\includegraphics[width=\linewidth]{{{Figure1}}}
\caption{\textbf{Graphene nanoribbon deflectometer.} \textbf{a}, Schematic of a graphene ribbon suspended between two gold contacts in aqueous solution. The water/ions are omitted and the deformed ribbon (green) has an exaggerated upward deflection, both for visual clarity. \textbf{b}, Hydrated ions and molecules can deflect the graphene, e.g., by binding to a functional group at the ribbon edge. \textbf{c}, In an alternative setup, the ribbon can cover an underlying window (e.g., in a silicon nitride membrane) and have a nanoscale pore, where hydrated ions or molecules (DNA, proteins, etc.)\ will impose a force on the pore edge during translocation. In either setup, flow can be driven by pressure or an applied bias.  \textbf{d}, A periodic ($10$~ns on, $10$~ns off) 100 pN force creates a series of deflected and undeflected events, where the dark blue line shows the moving average of the deflection (over $5$~ns)  -- that is, the average $z$-displacement of the carbon atoms in the suspended region of the ribbon relative to the edges at the gold contacts. The light blue shaded region represents one standard deviation of the atom positions propagated through the moving average. Thresholding gives the green (red) lines for the deflected (undeflected) events. The vertical placement of these lines has no significance other than to indicate the ``on/off'' deflection state. See the Methods for details. 
\label{fig:expdiagram}}
\end{figure*}

Novel sensing platforms and technologies often make use of transduction, i.e., they convert one input signal into another more easily detectable or transferrable signal. 
This is the enabling ingredient in techniques from piezoelectric actuation to alternative computing and communication paradigms~\cite{safari2008piezoelectric, rakher2010quantum,nguyen2012piezoelectric, ong2012engineered, liang2012electromechanical, manzeli2015piezoresistivity}, and even new biomedical sensors~\cite{yu_needle-shaped_2018}. We examine the transduction of mechanical strain -- a deflection induced by a small force -- into an electronic signal through a suspended graphene nanoribbon (GNR), Fig.~\ref{fig:expdiagram}a, as a route to sensing rapid and weak (bio)molecular events in solution at finite temperature.

When molecules flow by, e.g., a functionalized GNR edge, transient binding events will deflect the ribbon, see Fig.~\ref{fig:expdiagram}b,c. This, in turn, weakens the electronic couplings in the graphene according to the Hamiltonian 
\begin{equation} \label{eq:exphop}
\qh = \sum_{\< i j \>} v_0 e^{-l/\lambda} \cid{i} \ci{j} ,
\end{equation}
That is, the deflection gives an exponential decrease of the hopping energies, $v_0 e^{-l/\lambda}$, where $v_0 \approx-3$~eV is the hopping energy between $p_z$ orbitals in pristine graphene with creation (annihilation) operators $\cid{i}$ ($\ci{i}$) at lattice site $i$, $l$ is the carbon-carbon bond stretching, and $\lambda \approx 0.047$~nm is the length scale characterizing the decay (see the Methods and Ref.~\onlinecite{cosma2014strain}). That strain further opens up the band gap in GNRs (or nanotubes) in vacuum and on a substrate is well known~\cite{minot2003tuning, isacsson2011nanomechanical, cosma2014strain}. Here, we show that even in a room-temperature solution environment this opening is efficiently detectable above the noise, which will enable the use of GNR deflection for detecting molecular scale processes or even DNA sequencing~\cite{paulechka2016nucleobase, smolyanitsky2016mos2}.

\begin{figure*}
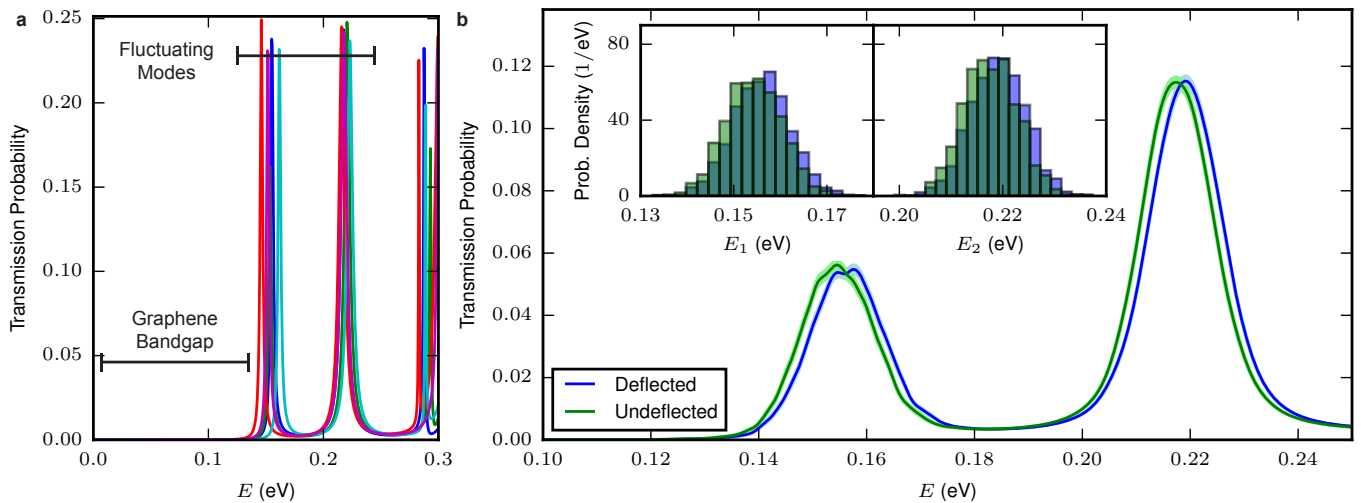

\includegraphics[width=\linewidth]{{{Figure2}}}
\caption{\textbf{Transmission probability through a graphene deflectometer.} \textbf{a}, The transmission probability for a random selection of instantaneous structures of the suspended graphene. As the structure changes, the electronic states shift in energy, although their height stays more or less the same (e.g., the contact resistance between the gold and graphene changes little). \textbf{b}, The average transmission probability for deflected and undeflected events using the division in Fig.~\ref{fig:expdiagram}d. The width of these average peaks are order $k_B T$ at room temperature. The blue and green shaded regions represent plus/minus one block standard error. The deflected state has a transmission peak shifted to higher energy. That is, the bandgap opens further, consistent with strain in a pristine nanoribbon in vacuum at zero temperature~\cite{isacsson2011nanomechanical, cosma2014strain}. We note that the dual maximum in the first peak of the deflected state is potentially real and a reflection of the fact that the ribbon is deflected on one edge rather than uniformly across the width. The energetic shift in current-carrying modes enables the detection of the deflection events, as the magnitude of the current will change. A shift in the Fermi level via liquid gating~\cite{ang_solution-gated_2008, traversi_detecting_2013,heerema_probing_2018} or an appropriately configured side gate 
can optimize sensitivity to this peak shift. Changing the magnitude of the bias also can help control the detectability of this signal. \textbf{Inset:} In any individual structure, 
there are transmitting modes at energies $E_1$ and $E_2$. The histograms of these peak positions show the shift in energy due to deflection, a shift that can be distinguished with sufficient sampling. The relative error in the binned data is in the Supplemental Information. See the Methods for details. 
\label{fig:optdiagram}}
\end{figure*}

Regardless of its origin, a force on one or a few carbon atoms in the GNR will result in its local deflection. Using all-atom molecular dynamics (see the Methods), we examine deflection by 100 pN forces, which is comparable with critical force to break hybridized DNA bases at nanosecond timescales~\cite{paulechka2016nucleobase}. 
Figure~\ref{fig:expdiagram}d shows a periodic force ($10$~ns on, $10$~ns off) applied to one carbon atom at the ribbon edge, giving rise to a series of deflected and undeflected events over the background noise due to thermal fluctuations. Figure~\ref{fig:optdiagram}a shows the electron transmission probability through a few example, instantaneous graphene structures. On top of the bandgap already present in a (slightly) strained ribbon of finite width, there are fluctuations of the states' energies. These fluctuations will be of order $k_B T$, where $k_B$ is Boltzmann's constant and $T$ is the temperature (here, room temperature). Expanding the exponent in the Hamiltonian, Eq.~\eqref{eq:exphop}, we see that the fluctuations in the coupling constants -- averaged over the whole structure -- are $v_0 \avg{l^2/\lambda^2}$, which, in thermal equilibrium, gives $v_0 k_B T/\kappa \lambda^2$ with $\kappa \approx 4360$ eV/nm$^2$ the carbon-carbon spring constant. At room temperature, this estimate is approximately 8 meV and will increase linearly with $T$. The thermal noise is thus directly imprinted on the probability for electrons to transmit through the ribbon.

In fact, the average transmission functions, Fig.~\ref{fig:optdiagram}b, seem to follow a Voigt profile commonly seen in solution/gas-phase spectroscopy and diffraction~\cite{wertheim_determination_1974}. Just as in those areas, transmission has two sources of broadening: One is the standard Lorentzian form (seen in Fig.~\ref{fig:optdiagram}a for the transmission through individual, instantaneous structures) due to coupling of the ribbon's electronic states to the electrodes. The other plays the role of thermal doppler broadening in spectroscopy: The thermal motions of the carbon atoms in graphene are a network of random bond strengths, giving rise to an effective parasitic ``random walk'' of the transmission peak position. However, the current does not yield a spectroscopic measurement of the transmission function due to both the finite bias and the thermal broadening of the electrons. As well, the average transmission actually follows a {\em generalized} Voigt form due to the presence of multiple neighboring transmission peaks. We will return to these points below. A Gaussian best fit for the average transmission peaks yields a full-width at half maximum of $(17 \pm 1)$~meV for all four peaks, i.e., a standard deviation of $\sigma = (7 \pm 0.5)$~meV, in line with the estimate above. 

The average transmission also shows that the current-carrying modes shift to higher energy when the GNR is deflected (Fig.~\ref{fig:optdiagram}b). We can estimate this shift directly from Eq.~\eqref{eq:exphop}: For a $0.2$~nm deflection of a $15$~nm long suspended region, the carbon-carbon bond stretch should be about $0.09$~pm on the deflected edge, or about half that on average across the width of the ribbon. Taking $l \approx 0.04$~pm, the change in electronic coupling is $\delta v \approx 2.7$~meV, which is approximately the shift shown in Fig.~\ref{fig:optdiagram}b [from a Gaussian best fit, one obtains $(1.6 \pm 0.5)$~meV]. Even with this small shift (compared to the thermal broadening), the strain-induced change in the current is efficiently detectable, as we will show.

We want to identify the deflection of -- or force on -- the graphene nanoribbon from the electronic current. This can be either in a continuous or discrete scenario, i.e., determining the magnitude of deflection/force or detecting a molecular binding event. In either case, detection requires separating the signal from intrinsic and extrinsic noise. To optimize detection, we want to minimize the quantity
\begin{equation}  \label{eq:OptFun}
\overbrace{\,\,\sigma_I / \chi \,\,}
         ^{\text{Continuous}}
\Leftrightarrow
\overbrace{\sigma_I / \Delta_I }
         ^{\text{\,\,\,\,Discrete\,\,\,\,}}.
\end{equation}
Here, the current noise, $\sigma_I$, divided by the susceptibility, $\chi$, of the current to deflection gives the appropriate measure of detectability (or, in the discrete case, divided by the absolute change in current, $\Delta_I=\chi \Delta_d$, where $\Delta_d$ is the change in deflection during the event). We consider $\sigma_I$ only from intrinsic sources and, to deal with extrinsic sources, we will also seek to maximize $\Delta_I$, as this gives a robust way to overcome noise due to the readout electronics. 

\begin{figure*}
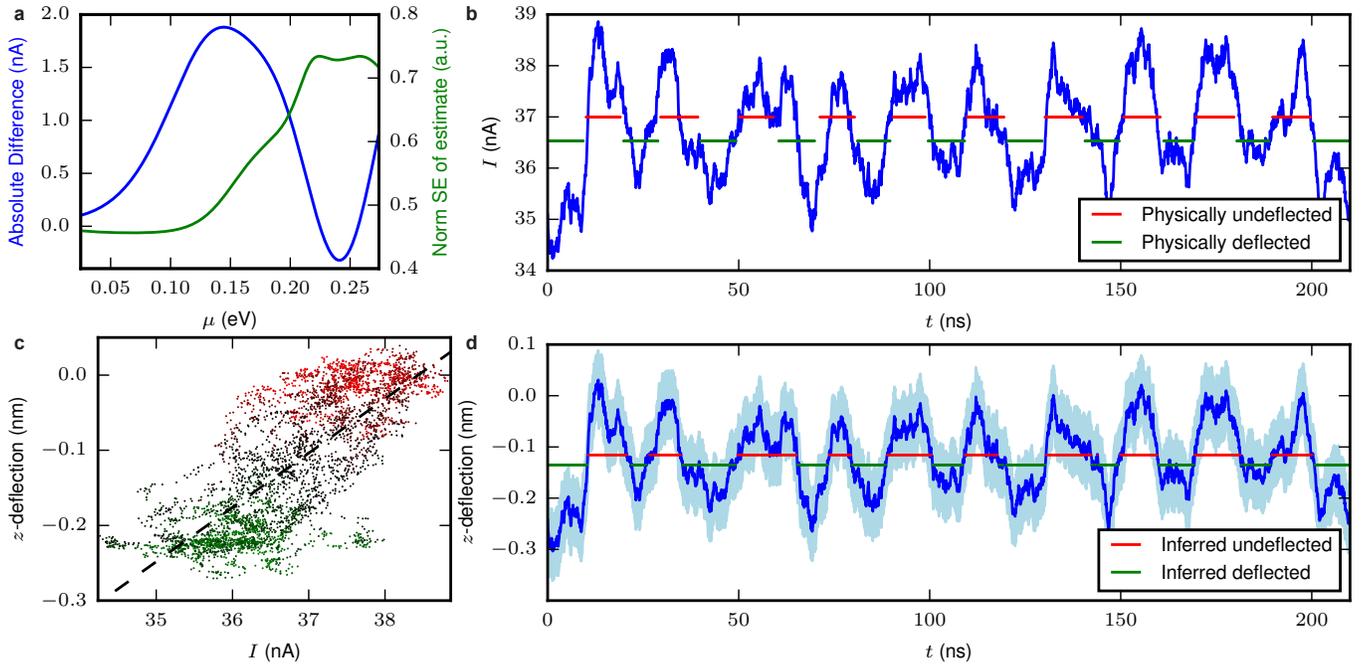

\includegraphics[width=\linewidth]{{{Figure3}}}
\caption{\textbf{Electronic current and deflection.} \textbf{a}, Absolute difference in the current at a 50 mV bias and the standard error of the estimate versus the Fermi level $\mu$. Although the $T(E)$ are nearly identical (Fig.~\ref{fig:optdiagram}), their absolute difference displays maxima at certain locations of the Fermi level. The error as propagated from the displacement and transmission functions is several orders of magnitude lower. It is advantageous to tune $\mu$ to give a large magnitude of absolute difference (i.e., to be above other external sources of noise) and a small error for inferring the deflection via the current (this error flattens at low energy due to the reduction of internal current fluctuations simultaneously with the reduction in the current magnitude). \textbf{b}, Electronic current as a function of time, as well as the on/off representation of the physical deflection from Fig.~\ref{fig:expdiagram}b. \textbf{c}, Scatter plot of the measured deflection and electronic current at $\mu=125$~meV, with the dashed line showing the least squares regression and the red-black-green color coding determine by a linear mixing of colors (red in the middle of when the force is applied, green when the force is not). The density of data points is highest for the deflected state ($z \approx 0.23$~nm, green) and the undeflected state ($z \approx 0$~nm, red). In between these two extremes there are structures transitioning from undeflected to deflected (or vice versa) just after a force is applied (released), as shown by black points. These structures are therefore sampled less (for instance, one sees a lone trajectory with a high current for a large range of $z$-deflection). \textbf{d}, Inferred deflection from the electronic current using the parameters of the linear regression and the deflection using the same thresholds as Fig.~\ref{fig:expdiagram}b. Approximately $76$~\% of the deflected and undeflected configurations are correctly represented in the current. The shaded region represents the error from the regression. 
\label{fig:reconstr}}
\end{figure*}

There are numerous potential methods to optimize these quantities, such as changing the baseline strain on the graphene or modifying other conditions (pH, temperature, gas-phase or vacuum instead of solution, GNR dimensions, etc.). We will consider the application of a uniform shift in Fermi level $\mu$ due to changing the pH~\cite{ang_solution-gated_2008,traversi_detecting_2013,heerema_probing_2018} or by a voltage gate. Figure~\ref{fig:reconstr}a shows the absolute current difference from the undeflected to deflected state, as well as the standard error of the estimate, which is just $\sigma_I / \chi$. We see that the optimal operating point is around $\mu=125$ meV, giving both a large current change upon deflection and small ``single shot'' expected error. That both conditions can almost simultaneously hold is not surprising, since increasing $\Delta_I$ decreases Eq.~\eqref{eq:OptFun}, all other effects being equal. 

We expect that the optimal operating regime should ``sit'' on a peak shoulder, as this is where the shift in the peak gives the maximal absolute change. 
At $\mu=125$ meV, the operation is near the first peak shoulder. However, we will see that due to the multiple sources of thermal broadening (of the electrons in the reservoirs and of the transmitting states), the optimal detection point is more strongly influenced by the second peak, the one at approximately 220 meV. To show this, let's examine the mean current under a deflection $d$, 
\begin{equation} \label{eq:meancurrent}
I_d = \frac{2 e}{h}\int_{-\infty}^{\infty} dE \left[f_\ql(E)-f_\qr(E)\right] \avg{T_d(E)},
\end{equation}
where $I_d=\chi (d-\avg{d})+\avg{I}$, $e$ is the electron charge, $h$ is Planck's constant, $f_{\ql (\qr)}$ are the Fermi-Dirac distributions in the left (right) reservoirs, and $\avg{\cdot}$ gives the average of the indicated quantity. Central to this expression is the average transmission function $\avg{T_d(E)}$ under deflection $d$. In the Methods, we show that a {\em single} Voigt profile for electrons moving through a fluctuating structure gives rise to the current
\begin{equation} \label{eq:Voigt}
A e^{\frac{\left(w/2\right)^2-\ue^2}{2 \sigma^2}} \cos \left[ \frac{w \, \ue}{2 \sigma^2} \right] \mathrm{Erfc} \left[ \frac{ w/2 - \im \ue}{\sqrt{2} \sigma} \right], 
\end{equation}
where $A$ is a prefactor and $w$ and $\ue$ are the Lorentzian peak width and offset from the Fermi level. The total thermal broadening $\sigma=\sqrt{\sigma_{FD}^2+\sigma_V^2}$ is due to both the electronic smearing, $\sigma_{FD}$, and transmission peak broadening, $\sigma_V \approx 7$ meV. The quantity  $2 \sigma_{FD} \sqrt{2 \ln 2} = 2 k_B T \arccos \left[ 2+\cosh (e V/2 k_B T) \right]$ is the full-width at half maximum of the finite temperature bias window [$f_\ql(E)-f_\qr(E)$]. 

The maximum change in the current under deflection, i.e., when the peak shifts (i.e., $\ue \to \ue+\delta v$), gives the most detectable signal. 
This occurs at (see Methods) $\ue \approx \pm \sigma$ (or $\mu \approx E_p \pm \sigma$, where $E_p$ is the peak position). For a $V=50$ mV bias at room temperature (and using $\sigma_V=7$ meV), this is approximately 40 meV away from the peak maximum. This is too low if measured from the first peak and too high from the second peak. However, when both peaks are considered jointly, the maximum change comes at $\mu \approx 135$ meV, in line with the full numerical results. At the maximum, about 2 nA changes occur. However, this drops rapidly, as $\exp \left( -\ue^2 \right/2 \sigma^2)$, away from the maximum.

We can also optimize inference, or how well we can determine the deflection state, by minimizing Eq.~\eqref{eq:OptFun} (either quantity will do, since they are related by a constant factor). Assuming that the current fluctuations are proportional to the current magnitude (which is the case away from the peak, see the Supplemental Information), we can minimize Eq.~\eqref{eq:OptFun} by maximizing the relative difference in the current where we choose the larger of the two currents (the undeflected current) to normalize the expression since it sets the largest noise. This gives the minimum in Eq.~\eqref{eq:OptFun} occurring at 
\begin{equation}
\ue = \sigma \sqrt{\ln \left[ \frac{8 \pi}{w^2} \right]}.
\end{equation}
Again, for a $V=50$ mV bias at room temperature (and using $\sigma_V=7$ meV), this is a shift of about 120 meV from the peak, much too low to result from the first peak, but quantitatively in agreement for a shift from the second peak. The presence of the first peak does, however, result in a much more slowly varying error in estimation for $\mu < 100$ meV, which is qualitatively in agreement with Fig.~\ref{fig:reconstr}a. While the maximum in the absolute change in current can be predicted from simpler approximations (e.g., a Gaussian or Lorentzian with additional broadening both predict the location of the maximum), the features in the behavior of the error estimation require recognizing that the average transmission is a Gaussian ``bulk'' and an algebraic tail (i.e., Voigt). 
 
The current at $\mu=125$ meV is shown in Fig.~\ref{fig:reconstr}b. It clearly follows the deflected and undeflected states. The magnitude of the current is $\approx 36$~nA, with a change of $\approx 2$~nA, consistent with the analytical estimate. The accuracy in detecting the on/off deflection state is about 80~\% when comparing the fidelity of the actual on/off time trace to that inferred from the current. In addition to detecting events, we can also extract the ``instantaneous'' deflection: Figure~\ref{fig:reconstr}c shows a linear regression between the filtered current and deflection, where the separation of the deflected and undeflected states is visible. Figure~\ref{fig:reconstr}d shows the reconstruction of the deflection using the electronic current. While more noisy than just detecting the on/off deflection, it is still accurate.

As is visually apparent from Fig.~\ref{fig:reconstr}b, we are near the detection limit: The fluctuations in the current over the interior of a 10 ns event (i.e., not in a transition state between undeflected and deflected) are about 1 nA. The separation between current levels of the deflected and undeflected states is about 2 nA (for the 100 pN force on the $15$~nm by $5$~nm ribbon edge giving a $\approx 0.2$ nm deflection). We can thus estimate the intrinsic physical limits of a graphene deflectometer. We have 100 pN forces at 200 MHz, which gives 7 fN$/\sqrt{\mathrm{Hz}}$, which we expect to be slightly larger than the intrinsic sensitivity limit. Events shorter than 10 ns or with forces (deflection) smaller than 100 pN (0.2 nm) will start to significantly degrade their detectability. However, longer events or a larger sampling time can resolve smaller strains/forces in a straightforward manner. Noise from readout electronics and environmental electrostatic fluctuations will diminish the sensitivity (in cases where electrostatic changes occur with the process to be detected, a sufficiently long, e.g., functional group needs to be used to keep a separation larger than the Debye length). Current measurements on GNRs~\cite{traversi_detecting_2013} show that the signals here are just below present-day sensitivity limits and further improvements~\cite{kim2013noise} may allow the integrated device to reach the intrinsic force sensitivity. 

Our results, including the simple analytical estimates and direct demonstration of the strain-current relationship in solution, show that graphene deflectometry is a promising route to sensing molecules and interactions in their native environment. Optimization of the detection via gating (or doping) and the bias brings its operation into realm of many molecular-scale interactions: It can sense forces less than 100 pN acting locally -- on the scale of the forces imparted by DNA base pair binding to functional groups~\cite{paulechka2016nucleobase, smolyanitsky2016mos2} --  and for a duration of less than 10 ns. These findings should open up the field of graphene deflectometry for sensing a broad range of processes, such as molecular binding and protein unfolding, as well as many more solution-phase phenomena.

\section*{Acknowledgments}

We thank K. A. Velizhanin for helpful discussions and S. Sahu for Fig.~\ref{fig:expdiagram}c. Daniel Gruss acknowledges support under the Cooperative Research Agreement between the University of Maryland and the National Institute of Standards and Technology Center for Nanoscale Science and Technology, Award 70NANB14H209, through the University of Maryland. Alex Smolyanitsky gratefully acknowledges support from the Materials Genome Initiative.

\section*{Contributions}

M.Z. and A.S. proposed the project. D.G. performed the electronic transport calculations and solved the inverse problem. A.S. devised and performed the MD simulations. M.Z. suggested and developed the theory of detection, optimization, and noise. All authors wrote the manuscript.

\section*{Additional information}

Correspondence and requests for material should be addressed to A.S. (alex.smolyanitsky@nist.gov) or M.Z. (michael.zwolak@nist.gov).

\section*{Competing Interests}

The authors declare no competing financial interests.

\section*{Methods}

\subsection{Molecular dynamics simulation of a GNR}

\textbf{Molecular dynamics simulation of graphene ribbon.} We performed all-atom molecular dynamics (MD) simulations of the graphene nanoribbon (GNR) immersed in water using the GROMACS 5.1.2 package~\cite{van2005gromacs}. The MD model of the GNR is based on the previously developed~\cite{smolyanitsky2014molecular} and employed~\cite{smolyanitsky2014molecular, smolyanitsky2015effects, paulechka2016nucleobase} bond-order informed harmonic potential, implemented within the OPLS-AA forcefield~\cite{jorgensen1988opls, jorgensen1996development}. The device is immersed in a rectangular container filled with explicit water molecules, using the TIP4P model~\cite{horn2004development, abascal2005potential}. Prior to the production MD simulations, all systems underwent NPT relaxation at $T = 300$~K and $p = 0.1$~MPa. All production simulations were in an NVT ensemble at $T = 300$~K, maintained by a velocity-rescaling thermostat~\cite{bussi2007canonical} with a time constant of $0.1$~ps. The graphene is slightly prestrained to increase the bandgap and provide a restoring force. Out-of-plane deflections of the GNR were due to a constant $100$~pN force in the $z$-direction, applied to a GNR carbon atom at $(w, L/2)$, where $w$ and $L$ are the GNR's width and length, respectively. The deflecting force was on a single carbon atom at the edge rather than the center, $(w/2, L/2)$, of the GNR to represent the ``side arrangement'' of a DNA sensor~\cite{smolyanitsky2016mos2}, which does not require a nanopore. The deflection is measured directly from the graphene coordinates by examining the average position of the carbon atoms in the center as compared to those along the outside edges. It should be noted that this arrangement implies a force detection only when the force imposes a structural change on the GNR.

\subsection{Electronic transport through the GNR}

\vspace{1em}

\noindent \textbf{Electronic transport calculations.} The total current through the graphene nanoribbon is found via~\cite{jauho1994time, gruss2016}
\begin{equation} \label{eq:totalcurr}
I = \frac{2e}{h}\int_{-\infty}^{\infty} dE \left[f_\ql(E)-f_\qr(E)\right] T(E)
\end{equation}
where the factor of 2 is for spin, $f_\ql$ and $f_\qr$ are the Fermi-Dirac distributions in the left and right reservoirs respectively, and the transmission function $T(E)$ is
\begin{equation}
T(E) = \tr \left[\bGa^{\ql}(E)\bG^{r}(E) \bGa^{\qr}(E)\bG^{a}(E)\right].
\end{equation}
The coupling to the reservoirs is $\bGa^{\ql(\qr)} = -2 \Im \bS^{\ql(\qr)}$ and the Green's functions $\bG$ are
\begin{equation}
\bG^r (E) = \frac{1}{E - \qh - \bS^\ql (E) - \bS^\qr (E)} ,
\end{equation}
where $\qh$ is the Hamiltonian and the self-energies $\bS^{\ql(\qr)}$ determine the coupling to the external reservoirs on the left (right) side of the system. These are set at a constant relaxation parameter, $\gamma$, which determines the rate of relaxation of the gold electronic states to thermal equilibrium in the absence of the graphene. In particular, $\gamma$ must be set to ensure that the current correctly reflects the intrinsic conductance of the ribbon in the presence of the gold contacts~\cite{gruss2016, elenewski2017master, gruss2017relaxation} (see the Supplemental Information for more details).

The ends of the suspended graphene are coupled to an additional $12$ hexagons ($\approx 2.5$~nm) of pristine graphene above three layers of a gold (111) surface. The applied potential is across the two gold substrates. The gold atoms have a $-3$~eV hopping energy between all adjacent atoms and an onsite energy of $-1.6$~eV, representing the $6s$-conduction band~\cite{wahiduzzaman2013dftb}. The parameter $\lambda$ in Eq.~\eqref{eq:exphop} is extracted from $\exp [\eta_0 (l/r - 1)]$, where $r=0.142$~nm, and $\eta_0 \approx -3$ (see Ref.~\onlinecite{cosma2014strain}). The carbon-gold coupling has a similar form with $r=0.331$~nm and equilibrium hopping of $-1.5$~eV. 

\subsection{Optimization and the Voigt Profile}

\vspace{1em}

\noindent \textbf{Optimization and the Voigt Profile.} In the scenario in Fig.~\ref{fig:expdiagram}d, detection amounts to inferring the on/off deflection state (or determining the magnitude of the applied force for continuous detection). A deflection, $d$, will result in some probability distribution $P_d(I)$ of output currents, e.g., $P_d(I)\propto \exp\left[ (I-I_d)^2/2\sigma_I^2 \right]$, where $I_d=\chi (d-\avg{d})+\avg{I}$ and $\chi$ is the susceptibility of the current to the deflection. To infer whether the ribbon is deflected, we only need to know whether the current is above some threshold ($\avg{I}$ in this case where the on/off states are equally probable). If so, the undeflected state is most likely. If not, the deflected state is most likely. The error of incorrectly identifying a the deflected (undeflected) state is just the integral of $P(I)$ over $I$ above (below) the threshold, which yields an expected error probability of $\mathrm{Erfc}\left[\Delta_I / 2 \sqrt{2} \sigma_I \right]$. For optimal detection, therefore, we want to minimize 
\begin{equation} \label{eq:OptFun2}
\frac{\sigma_I}{\Delta_I},
\end{equation} 
where $\Delta_I$ is the separation of the current between the two deflection states. Minimizing this quantify ensures that the fluctuations in current (within a given deflection state) are the smallest possible compared to the change in mean current upon deflection, giving optimal inference of the deflection state from the current (in the absence of external sources of noise). 

For continuous detection, one wants to infer the deflection (or force) from the value of the current. Transforming the distribution, $P_d(I)$, to one over the deflection gives $P_I (d) \propto \exp\left[ (d-d_I)^2/2\sigma_d^2 \right]$ with $\sigma_d=\sigma_I/\chi$ and $d_I=(I-\avg{I})/\chi+\avg{d}$. With this distribution, the standard error of the estimate is
\begin{equation}
\frac{\sigma_I}{\chi}.
\end{equation}
In the numerical results, we compute this directly from the data, see Eq. (S6), and do not assume a Gaussian distribution. 

The current through a single structure is given by Eq.~\eqref{eq:totalcurr}. The main part of each transmission peak $i$ through a single instantaneous structure is a Lorenztian,
\begin{equation}
T_i(E)= p_i \frac{(w_i/2)^2}{(E-E_i)^2+(w_i/2)^2},
\end{equation}
where $p_i$ is the peak height, $w_i$ the peak width, and $E_i$ the peak position. We caution that the actual average transmission functions, while Lorentzian in the bulk of the peak, has a different tail. This is due to the presence of nearby neighboring transmission peaks (see the Supplemental Information). The resulting profile, however, captures all the features we see in Fig.~\ref{fig:reconstr}a. 

All parameters of the Lorenztian have some distribution due to the fluctuations of the GNR. The fluctuations in the peak position is the most important source of noise for detecting the shift, as the other parameters all follow similar distributions, see Fig. S3. Thus, we consider the Voigt distribution
\begin{equation} \label{eq:AvgT}
\avg{T_d(E)} = \int dE_i \frac{1}{\sqrt{2 \pi \sigma_V^2}} e^{-(E_i-E_d)^2/2 \sigma_V^2} T_i(E) 
\end{equation}
where $E_d$ is the average peak position of the deflection state $d$. Employing Eq.~\eqref{eq:AvgT} in Eq.~\eqref{eq:meancurrent} gives the mean current. This total expression can not be integrated analytically. However, so long as the Fermi level is near (within about 100 meV) to the peak position, one can give an accurate approximation to the bias window,
\begin{equation}
\left[f_\ql(E)-f_\qr(E)\right] \approx \tanh \left[ \frac{V}{4 k_B T} \right] e^{-(E-\mu)^2/2 \sigma_{FD}^2},
\end{equation}
which has the right window height, $\tanh \left[ V/ 4 k_B T \right]$, and the right window width, $\sigma_{FD}$, at bias V and temperature $T$. The tail decays too fast compared to the exact bias window, limiting the accuracy of this expression when the separation of the peak and the Fermi level is greater than about 100 meV. With this approximation, we get the expression in Eq.~\eqref{eq:Voigt}. The prefactor in that expression is $A= \tanh \left[ V/4 k_B T \right] \sigma_{FD} p \gamma \pi \cdot 2 e/h$. The expressions for the location of the maximum current change and for the minimal error estimation come from Eq.~\eqref{eq:Voigt} by expanding for small $\delta v$ and for small $w$ (both of which are much smaller than all other energy scales involved).




\end{document}


\author{Daniel Gruss}

\affiliation{Center for Nanoscale Science and Technology,
             National Institute of Standards and Technology,
             Gaithersburg, MD 20899}
\affiliation{Maryland Nanocenter, University of Maryland,
             College Park, MD 20742}

\author{Alex Smolyanitsky} 

\affiliation{Applied Chemicals and Materials Division,
             National Institute of Standards and Technology,
             Boulder, CO 80305}

\author{Michael Zwolak} 

\affiliation{Center for Nanoscale Science and Technology,
             National Institute of Standards and Technology,
             Gaithersburg, MD 20899}

\title{Graphene deflectometry for sensing molecular processes at the nanoscale -- Supplemental Information}


\maketitle

\tableofcontents

\clearpage

\renewcommand\thefigure{S\arabic{figure}} 
\renewcommand\theequation{S\arabic{equation}}
\renewcommand\thetable{S\arabic{table}}
\renewcommand\thesection{\arabic{section}}
\renewcommand\thesubsection{\thesection.\arabic{subsection}}

\section{Methods}

\subsection{Deflection}

Assuming that the GNR deflected at the center forms a triangle with its undeflected state, the central $z$-deflection is taken as twice the average $z$-coordinate of the suspended region relative to the restrained edges: $z_{\text{defl}} = 2 \< z_{\text{sus}} \> - 2 \< z_{\text{end}} \> $, where a carbon atom is included in $\< z_{\text{sus}} \>$ if $1$ nm $\leq z \leq  14$~nm and is in $\< z_{\text{end}} \>$ otherwise. This expression is the lowest order expansion of the length change of the nanoribbon assuming a right triangle. The fluctuation in this quantity is a combination of the standard deviations of each one: $\sigma_{\text{defl}} = 2 \sqrt{\sigma_{\text{sus}}^2 + \sigma_{\text{end}}^2}$. The high frequency noise is removed before  further analysis of the deflection (the current calculations use the unfiltered, instantaneous structures) by applying a moving average of $5$~ns (or 100 instantaneous structures), with the start and end structures wrapped to eliminate edge effects.

Partitioning into deflected, undeflected, and intermediate structures is done automatically with the filtered $z$-deflection and a threshold of $z_{\text{defl}} < -0.135$~nm for deflected structures and $z_{\text{defl}} > -0.1155$~nm for undeflected (set through a visual inspection). We then group these into sets of at least ten continuous structures with four structures removed from the boundary of each range to account for the uncertainty in the determining that boundary.

\subsection{Electronic transport}

For each $50$~ps step in the instantaneous, MD-simulated structure (without any filtering of the atomic positions), the transmission function, $T(E)$, is calculated directly, which gives the current
\begin{align} \label{eq:totalcurr}
I &= \frac{2e}{h}\int_{-\infty}^{\infty}dE \;
          \left[f_\ql(E)-f_\qr(E)\right]
          \times \tr \left[\bGa^{\ql}(E)\bG^{r}(E)
                            \bGa^{\qr}(E)\bG^{a}(E)\right] \notag \\
  &= \frac{2e}{h}\int_{-\infty}^{\infty}dE \;
          \left[f_\ql(E)-f_\qr(E)\right]
          \times T(E) ,
\end{align}
where $f_\ql$ ($f_\qr$) is the Fermi-Dirac distributions in the left (right) reservoir and $\bG$ is 
\begin{equation}
\bG^r (E) = \frac{1}{E - \qh - \bS^\ql (E) - \bS^\qr (E)} 
\end{equation}
with $\qh$ the Hamiltonian. The $\bS^{\ql(\qr)}$ terms determine the coupling to the external reservoirs on the left (right) side of the system and are set to a constant, $\gamma$. Here,  $\ql$ and $\qr$ are the left and right gold substrates. When $\ql$ and $\qr$ are independently diagonalized,
\begin{equation}
\bS^\ql_{ij} = - \frac{\im \gamma}{2} \sum_{k \in \ql} \delta_{ij} \delta_{ik} , 
\qquad 
\bS^\qr_{ij} = - \frac{\im \gamma}{2} \sum_{k \in \qr} \delta_{ij} \delta_{ik} ,
\end{equation}
where the spectral functions are $\bGa^{\ql(\qr)} (E) = - 2 \Im \bS^{\ql(\qr)} (E)$.

\begin{figure}
\includegraphics[width=\linewidth]{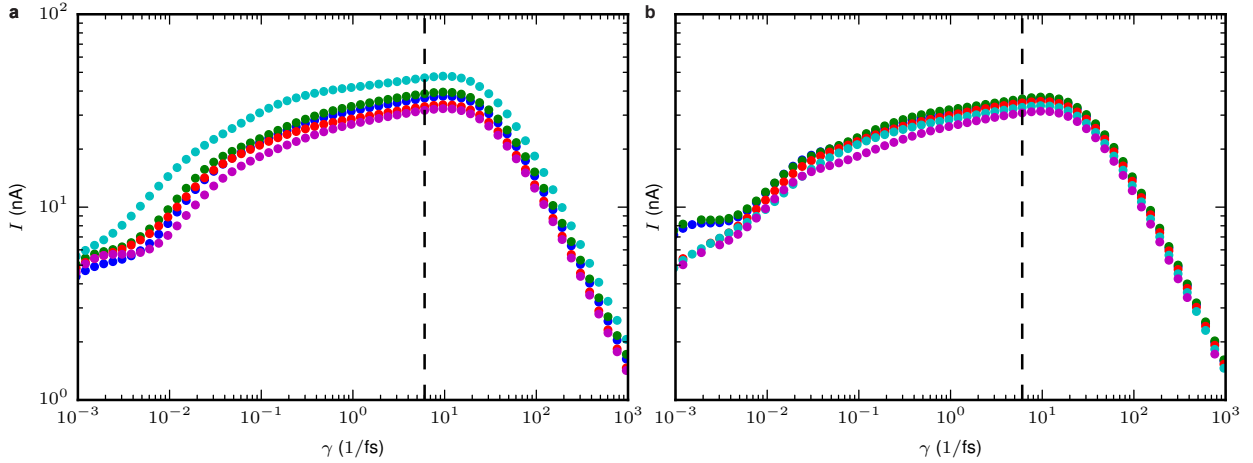}
\caption{\textbf{Current versus relaxation strength.} To correctly calculate the intrinsic current of the nanoribbon, the relaxation parameter, $\gamma$, should not be too small -- i.e., be the rate limiting process for the conductance -- nor too large -- causing localization in the reservoir regions. The plots show $I$ versus $\gamma$ for a sample subset of, \textbf{a}, deflected structures and, \textbf{b}, undeflected structures. In the main text we use the relaxation rate $\gamma \approx 6.05 /$fs (dashed, vertical line), which is within the central, quasi-plateau region. 
\label{fig:ivgamma}}
\end{figure}

Figure~\ref{fig:ivgamma} shows the electronic current as a function of $\gamma$. In the small- and large-$\gamma$ limits, the total current is limited by the relaxation rate. The intermediate regime determines the intrinsic conductance of the ribbon ``in isolation'': The region of interest does not have a conductance independent of how it is contacted. Thus, the intrinsic conductance is the conductance of the whole setup where $\gamma$-dependence has been eliminated~\cite{Gruss2016, elenewski2017master, gruss2017relaxation}. The sizes of the gold substrate and the value of $\gamma$ are chosen so that this dependence has been effectively removed but for which the computational cost of the simulation is still reasonable. 

\begin{figure}
\includegraphics[width=0.66\linewidth]{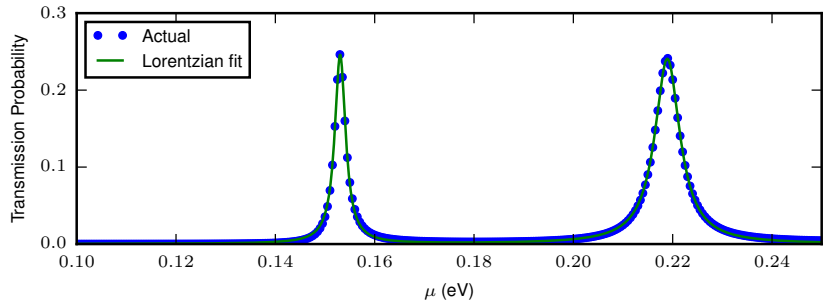}
\caption{\textbf{Lorentzian transmission peaks.} The transmission of the first two modes (those most relevant for detection) is given by two Lorentzian peaks, as is the case in nanoscale devices (individual transmitting modes are eigenstates weakly coupled to the electrodes). The resulting fitting parameters (peak position, width, and scale) can then be analyzed to determine the effect of deflection. \label{fig:curvefit}}
\end{figure}

Within the energy range of Fig. 2b of the main text, the transmission function, $T(E)$, has two Lorentzian peaks corresponding to broadened electronic modes, i.e., it has the form 
\begin{equation} \label{eq:twolorentz}
T(E) \approx \frac{p_1 (w_1/2)^2}{(E-E_1)^2 + (w_1/2)^2} + \frac{p_2 (w_2/2)^2}{(E-E_2)^2 + (w_2/2)^2} ,
\end{equation}
where $p_i$ is the peak height, $w_i$ is the width, and $E_i$ is the peak location of peak $i=1,2$. Figure~\ref{fig:curvefit} shows an example fit for a single, instantaneous structure.

\begin{figure}
\includegraphics[width=\linewidth]{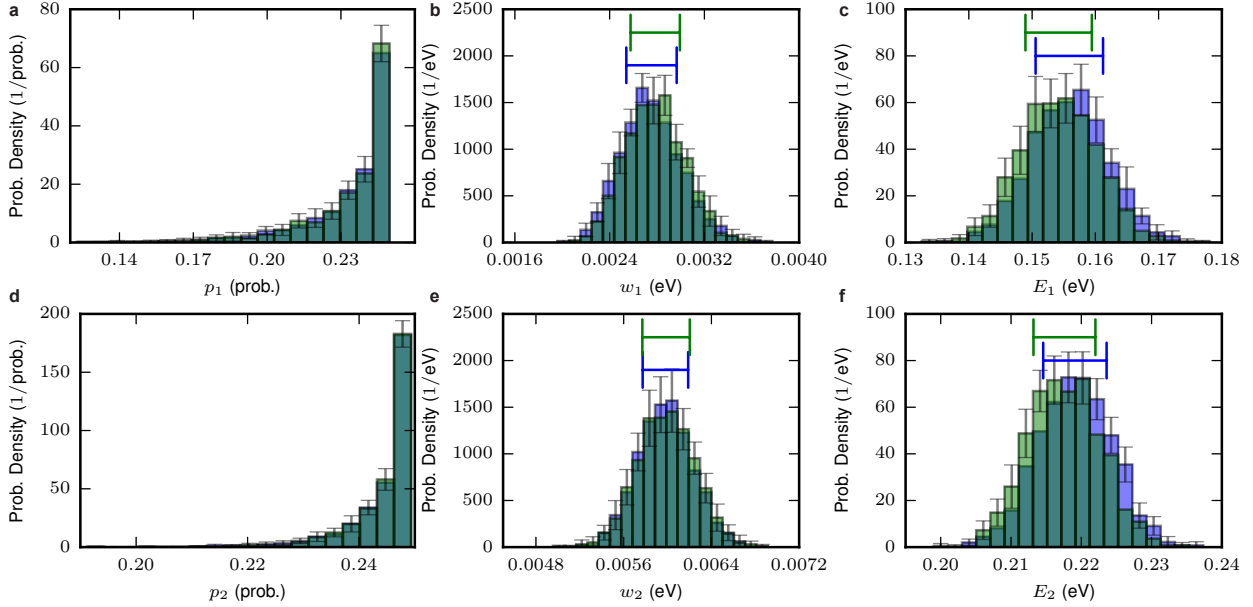}
\caption{
\textbf{Histograms of the Lorentizian parameters.} Probability density of the fit parameters in Eq.~\eqref{eq:twolorentz} for the deflected (blue) and undeflected (green) structures for both the first peak (\textbf{a}, \textbf{b}, \textbf{c}) and the second peak (\textbf{d}, \textbf{e}, \textbf{f}) of the transmission function. The overall peak height (\textbf{a}, \textbf{d}) and width (\textbf{b}, \textbf{e}) are mostly constant (as compared to their scale), while the peak positions (\textbf{c}, \textbf{f}) show a clear difference between deflected and undeflected states. The blue and green lines on each figure show the mean and the extent of one standard deviation of a Gaussian fit for the deflected and undeflected structures. The error bars are derived from the block standard error for each histogram bin (taking the maximum between the deflected and undeflected cases).
\label{fig:curvehist}}
\end{figure}

Figure~\ref{fig:curvehist} shows histograms for these parameters for the deflected and undeflected structures. The dominant effect is a shift in the energy when the nanoribbon undergoes deflection. This is consistent with a strain-induced shift of the electronic couplings. A change of contact resistance during deflection, for instance, would lower the peak height. The probability distributions are approximately Gaussian (although properly sampling the distribution for large shifts is computationally demanding), giving a thermal ensemble of individual Lorentzian peaks. In other words, the thermal noise and electronic couplings yield the Voigt profile discussed in the main text. Table~\ref{tbl:gaussfits} shows the parameters found from a non-linear fit of a Gaussian to the histograms for $w$ and $E$.

\begin{table}
\begin{tabular}{|c|c|c|c|c|}
\hline 
• & \multicolumn{2}{|c|}{Deflected} & \multicolumn{2}{|c|}{Undeflected}  \\ 
\hline 
• & Average & Standard Deviation & Average & Standard Deviation \\ 
\hline 
$w_1$~(meV) & $2.753 \pm 0.006$ & $0.256 \pm 0.005$ & $2.785 \pm 0.006$ & $0.251 \pm 0.005$ \\ 
\hline 
$w_2$~(meV) & $5.980 \pm 0.003$ & $0.254 \pm 0.003$ & $5.986 \pm 0.004$ & $0.263 \pm 0.004$ \\ 
\hline 
$E_1$~(meV) & $155.9 \pm 0.1$ & $6.14 \pm 0.09$ & $154.2 \pm 0.1$ & $6.1 \pm 0.1$ \\ 
\hline 
$E_2$~(meV) & $219.09 \pm 0.09$ & $5.23 \pm 0.07$ & $217.6 \pm 0.2$ & $5.1 \pm 0.1$ \\ 
\hline 
\end{tabular}
\caption{\textbf{Gaussian fit of histograms.} For each of the deflected and undeflected histograms in Fig.~\ref{fig:curvehist}(\textbf{b}, \textbf{c}, \textbf{e}, \textbf{f}), a non-linear fit to a Gaussian gives the mean and the standard deviation for the distribution. The corresponding errors are one standard deviation from the estimated covariance.}
\label{tbl:gaussfits}
\end{table}

\begin{figure}
\includegraphics[width=0.66\linewidth]{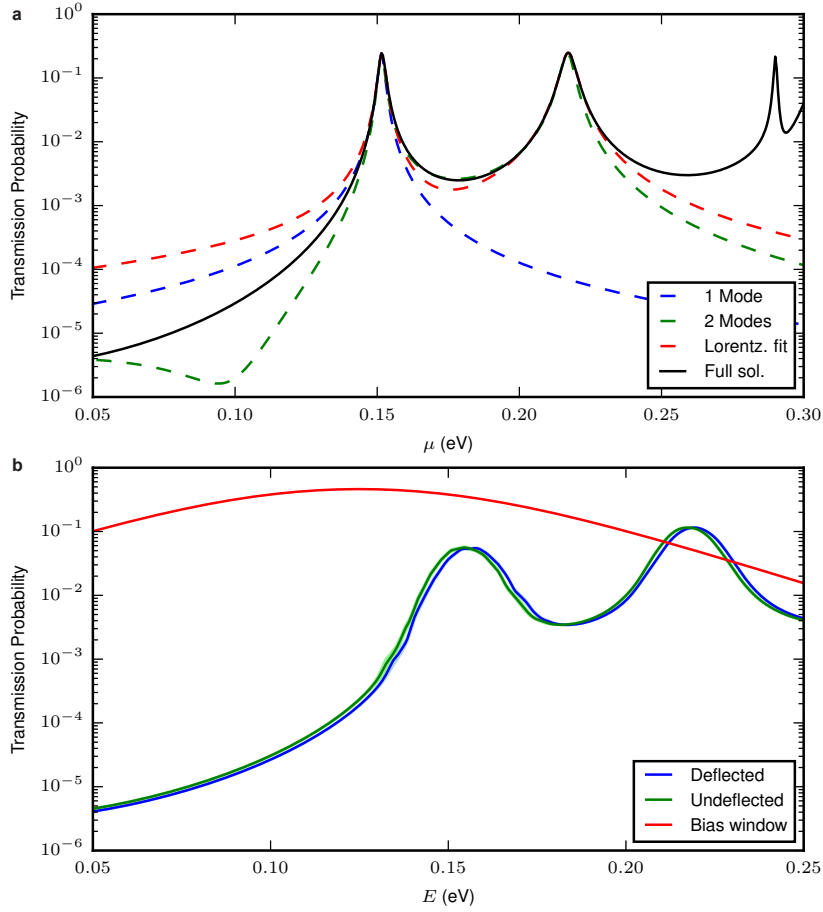}
\caption{\textbf{Tail properties of transmission function.} \textbf{a}, Examination of the transmission probability on a logarithmic scale shows that the exact solution (black line) has a different slope at low energy than the Lorentzian fit (red dashes). When $\qs$ is treated as a single mode around at that energy (blue dashes), it also displays an incorrect slope. Only when multiple modes are projected in aggregate (green dashes) does it show a differing slope, although other features appear as well indicating that additional modes are necessary to fully capture this tail. \textbf{b}, For the averaged transmission functions from the main text, the tails of the profiles decrease mostly in tandem rather than immediately converge. In addition, the bias window placement at the optimal value (red line) enhances the detection of the shift in the second peak.
\label{fig:tailprop}}
\end{figure}

The Lorentzian fits appear to show good agreement with the exact transmission function (and provide the basis for the subsequent use of the Voigt profile). However, it should be noted that the tail characteristics of the transmission function can differ from a true Lorentzian, as seen in Fig.~\ref{fig:tailprop}, due to the off-diagonal terms in the self-energy. This can be seen by calculating the transmission function using a subset of the important modes in the region of interest~\cite{gruss2017relaxation}. When the two main peaks are projected from the full solution, we find a transmission that is lower than both a pure Lorentzian and the full solution, meaning that higher order effects play a role in determining the structure of the transmission function away from the peak. To optimize the absolute difference between the deflected and undeflected currents (in order to maximize the measurable current), this difference plays a negligible role (as opposed to the maximum \emph{relative} difference, which depends on the tail structure).

\subsection{Numerical considerations and error quantification}

\begin{figure}
\includegraphics[width=0.5\linewidth]{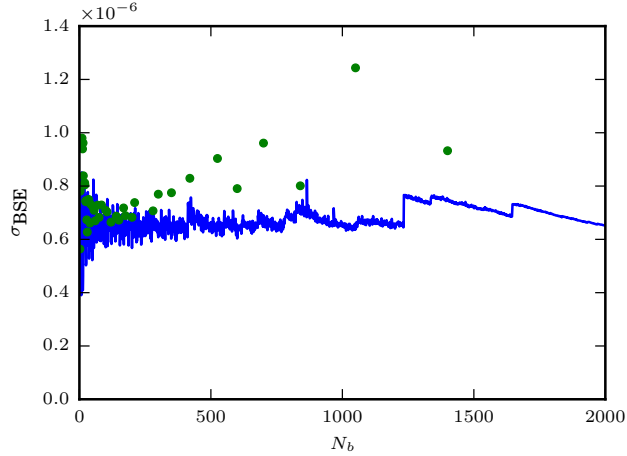}
\caption{\textbf{Block standard error.} The block standard error for the deflected structures using random sampling (blue line) and contiguous equally-sized blocks (green dots) of the current for an example snapshot. For error estimation purposes, the two methods are approximately equal, with the random sampling of structures being statisically independent. Away from either extreme of the block size, the error remains fairly constant. \label{fig:bse}}
\end{figure}

When a set of structures is treated in aggregate, the block standard error is~\cite{grossfield2009}
\begin{equation}
\sigma_\text{BSE} = \frac{\sqrt{\tau}}{\sqrt{T}} \sqrt{\frac{\sum_i \left( \< I_\tau \>_i - \< I_T \> \right)^2}{N_b - 1}} ,
\end{equation}
where $T$ is the total simulation time, $\tau$ is the length of time of a single block, $\< I_\tau \>$ is the average current within a block, $\< I_T \>$ is the total average current, and $N_b$ is the total number of blocks. Figure~\ref{fig:bse} shows the error as a function of block size, which is approximately constant at $N_b = 500$, which is the value we use for the error in the main text and SI plots. 

For the regression between the electronic current and the $z$-deflection, the error is the standard error of the estimate
\begin{equation}
\sigma_{\text{est}} = \sqrt{\frac{\sum (z_\text{defl} - z_\text{rec})^2}{N - 2}} ,
\end{equation}
where $z_\text{rec}$ is the reconstructed $z$-deflection and $N$ is the total number of instantaneous MD structures. In Fig. 3a of the main text, this quantity is normalized by the average total deflection, $\sum z_\text{defl} / N$.

\section{Currents in different operating regimes}

\begin{figure}
\includegraphics[width=\linewidth]{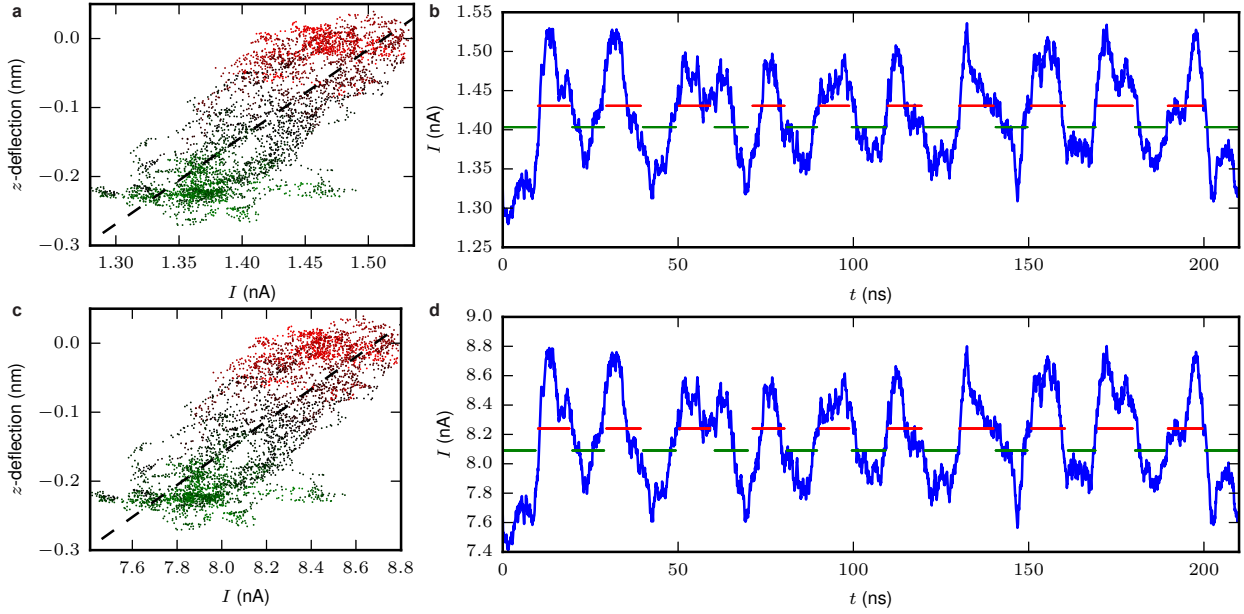}
\caption{\textbf{Inference of the deflection.} For a Fermi level of $\mu=25$~meV (\textbf{a}, \textbf{b}) and $\mu=75$~meV (\textbf{c}, \textbf{d}), the total current behavior is functionally similar to that seen in the mean text due to the same standard error, but at a much lower current magnitude. The scatter plots of the current and deflection (\textbf{a}, \textbf{c}), as well as the current versus time (\textbf{b}, \textbf{d}), show that the noise scales with the magnitude of the current. Nevertheless, if the physical measurement of a smaller current is possible at high speeds (i.e., that the readout noise is sufficiently low), the $z$-deflection can be determined with  $\approx 77$ \% of the total structures correctly identified as deflected or undeflected in both cases.
\label{fig:altcurr}}
\end{figure}

The main text uses a Fermi level of $\approx 125$~meV and a $50$~mV bias. However, since the bias window is large due to thermal broadening, there are potentially many suitable values for the Fermi level that will enable detection of deflection. Figure~\ref{fig:altcurr} shows the inferred deflection using an alternative value of the Fermi level with a similar gradient error but a lower absolute difference. For this particular system, it produces nearly the same result as the main text, but with a lower magnitude of the total current, which would be more difficult to measure in practice.

